\documentclass[traditabstract,preprint2]{aa}
\usepackage{txfonts}
\usepackage{natbib}
\bibpunct{(}{)}{;}{a}{}{,}
\usepackage[pdftex]{graphicx}
\usepackage{indentfirst} 
\usepackage{amssymb}

\usepackage{verbatim}

\begin{document}


\title{On vertical variations of gas flow in protoplanetary disks and their impact on the transport of solids}

\author{Emmanuel Jacquet\inst{1}}

\institute{Canadian Institute for Theoretical Astrophysics, University of Toronto, 60 St Georges Street, Toronto, ON M5S 3H8, Canada. \email{ejacquet@cita.utoronto.ca}}



\abstract{A major uncertainty in accretion disk theory is the nature and properties of gas turbulence, which drives transport in protoplanetary disks. 
 The commonly used viscous prescription for the Maxwell-Reynolds stress tensor gives rise to a meridional circulation where flow is outward near the midplane and inward away from it. This meridional circulation has been proposed as an explanation for the presence of high-temperature minerals (believed to be of inner solar system provenance) in comets. However, it has not been observed in simulations of magnetohydrodynamical (MHD) turbulence so far. 
In this study, we evaluate the extent to which the net transport of solids can be diagnostic of the existence of meridional circulation. To that end, we propose and motivate 
 a prescription for MHD turbulence which has the same free parameters as the viscous one. We compare the effects of both prescriptions on the radial transport of small solid particles and find that their net, vertically integrated radial flux is actually quite insensitive to the flow structure for a given vertical average of the turbulence parameter $\alpha$, which we explain. Given current uncertainties on disk turbulence, one-dimensional models are thus most appropriate to investigate radial transport of solids. A corollary is that the presence of high-temperature material in comets cannot be considered an unequivocal diagnostic of meridional circulation. In fact, we argue that outward transport in viscous disk models is more properly attributed to turbulent diffusion rather than to the mean flows of the gas. 
}

\titlerunning{Vertical variations of gas flow and solid dynamics}
\authorrunning{E. Jacquet}

\maketitle

\section{Introduction}
Extraterrestrial samples provide ample evidence for significant radial transport in the protoplanetary disk from which the solar system emerged. For instance, carbonaceous chondrites contain high-temperature materials such as refractory inclusions, believed to have formed close to the Sun in the first stages of disk building \citep{Wood2004}, chondrules, formed 1-3 Ma later \citep{Villeneuveetal2009}, as well as 
 aqueous alteration products indicating the presence of ice when these meteorites accreted. Yet more spectacularly, rare refractory objects and chondrule-like fragments have been identified in dust returned from comet Wild 2 \citep{Zolenskyetal2006, Bridgesetal2012}. 

 One way of transporting solids outward from the vicinity of the Sun is through bipolar outflows, as in the X-wind model of \citet{Shuetal2001} (see also \citealt{Hu2010}); however, whether solids can be present at and efficiently transported from the X point has been called into question by \citet{Deschetal2010}. More recently, scenarios have been proposed where transport is effected inside the disk by turbulent motions of the gas --- that same turbulence which would also drive accretion of the disk gas onto the central star \citep{BockeleeMorvanetal2002,Boss2004,Carballidoetal2005,JohansenKlahr2005,Ciesla2009,HughesArmitage2010,Jacquetetal2011a,YangCiesla2012}. In early stages, where the disk is massive, such turbulence could be due to gravitational instabilities \citep{Boss2004} whereas magnetohydrodynamical (MHD) turbulence powered by the magnetorotational instability (MRI; \citealt{BalbusHawley1998}) should dominate afterward. However, the MRI is expected to be suppressed over a significant range of heliocentric distances because of insufficient ionization \citep{Gammie1996}, so that in this region, referred to as the ``dead zone'', turbulence 
 should be reduced and have a more hydrodynamical character.

 While many studies have restricted attention to one-dimensional disk models, where only the vertically averaged turbulence parameter $\alpha$ had to be prescribed, some have attempted to calculate the transport of solids in two- or three-dimensional disk models \citep[e.g.][]{Hersantetal2001,Ciesla2009,Charnozetal2011,Ciesla2010vertical,Ciesla2011radial}. Going to 2D or 3D requires however more assumptions on the properties of turbulence to be made despite the considerable uncertainties of accretion disk theory and the diversity of potential turbulence drivers alluded to in the previous paragraph. A widespread prescription models turbulence as an effective viscosity in the averaged dynamical equations of the gas \citep[e.g.][]{TakeuchiLin2002}: the circulation resulting from this ``viscous'' prescription typically involves an outward flow around the disk midplane and inward flows in the ``atmosphere'', with the vertically integrated flow being inward (i.e. a positive mass accretion rate). This \textit{meridional circulation} was found by \citet{Ciesla2007} to improve outward transport of inner solar system material and hence account for the presence of high-temperature minerals in comets. However as yet, the meridional circulation is an essentially theoretical construct that has not yet been observed in numerical simulations of turbulent disks, in particular in the global MHD simulations of \citet{Fromangetal2011} and \citet{Flocketal2011}. As yet, the vertical structure of the gas flow is uncertain.

  The purpose of this article is to investigate, through analytical calculations, to what extent the net outward transport of small solids can be considered a diagnostic for meridional circulation. We propose and motivate a prescription for the vertical profile of the Maxwell-Reynolds stress tensor in the case of MHD turbulence and calculate the resulting mean flow of the gas. The prescription, inspired by previous numerical studies, formally depends on the same parameters as the standard ``viscous'' prescription, allowing a direct comparison. In particular, we compute the net (vertically integrated) radial flux of solids in both prescriptions. It is found that this net flux is weakly dependent on the prescription used, which will be explained, and that therefore meteoritic and cometary properties cannot be used as evidence for a specific vertical flow profile.

  This article is organized as follows: in Section \ref{Gas}, we introduce general notions on gas turbulence and the two prescriptions considered here. In Section \ref{Impact on the dynamics of solids}, we investigate and compare the dynamics of solids in both prescriptions. We discuss the results in Section \ref{Discussion} before concluding in Section \ref{Conclusion}.

\section{Modeling of gas turbulence}
\label{Gas}

In this section, we introduce 
 the modeling of gas turbulence. After having reviewed general equations, by way of establishing notation
, we introduce the two prescriptions of the Maxwell-Reynolds tensor investigated in this paper, namely the viscous and the MHD prescriptions
.

\subsection{Generalities}
\label{General equations}

The disk is described in a cylindrical coordinate system, with $R$ the heliocentric distance, $z$ the altitude above the midplane, and $\phi$ the azimuthal angle. 
 We note $\mathbf{u}$, $\rho$, $T$ and $P=\rho c_s^2$ the gas velocity, density, temperature and pressure, respectively, with $c_s$ 
the isothermal sound speed
. $v_K=\sqrt{GM_\odot/R}$ and $\Omega = v_K/R$ are the Keplerian linear and angular velocities, respectively.
 
  We treat the disk as vertically isothermal\footnote{Although this approximation breaks down in the surface layers of the disk \citep[e.g.][]{ChiangGoldreich1997}, this is of no concern here as this generally affects a small fraction of the surface density of the gas and an even smaller one of that of the (partly settled) solids.} so that vertical hydrostatic equilibrium implies the following density stratification:
\begin{equation}
\rho(R,z)=\frac{\Sigma(R)}{\sqrt{2\pi}H(R)}\exp{\left(-\frac{z^2}{2H(R)^2}\right)}
\label{density stratification}
\end{equation}
with the pressure scale height $H=c_s/\Omega$ and  the surface density $\Sigma\equiv\int_{-\infty}^{+\infty}\rho\mathrm{d}z$. 

  We assume the disk to be turbulent and assume axisymmetry in the sense that variations of any quantity $Q$ in the azimuthal direction may be treated as turbulent fluctuations about a mean $\overline{Q}$. We denote the Eulerian perturbations with $\delta Q\equiv Q-\overline{Q}$. 
For any quantity $Q$ and a weight function $w$, we also define the $w$-weighted vertical average of $Q$ as:
\begin{equation}
\langle Q\rangle_w\equiv\frac{\int_{-\infty}^{+\infty}Q(z)w(z)\mathrm{d}z}{\int_{-\infty}^{+\infty}w(z)\mathrm{d}z}.
\end{equation}

  Because of the pressure gradient and the departure from the equatorial plane for $z\neq 0$, $u_\phi$ is not exactly equal to the Keplerian velocity $v_K$. Indeed,  centrifugal balance yields:
\begin{eqnarray}
\overline{u_\phi}(R,z) & = & \sqrt{\frac{v_K^2}{\left(1+(z/R)^2\right)^{3/2}}+ \frac{1}{\rho}\frac{\partial P}{\partial\mathrm{ln}R}}\nonumber\\
& = & v_K+\frac{1}{2\rho(R,0)\Omega}\frac{\partial P(R,0)}{\partial R} + \frac{\Omega}{4}\frac{\partial\mathrm{ln}T}{\partial R}z^2 + o\left(\left(\frac{H}{R}\right)^2\right)
,
\label{subkeplerian}
\end{eqnarray}
where we have used $z, H\ll R$. Since, generally, $\partial P(R,0)/\partial R < 0$ and $\partial\mathrm{ln}T/\partial R <0$, the flow is subkeplerian 
 \citep[e.g.][]{Fromangetal2011, Flocketal2011}. However, the approximation $u_\phi\approx v_K$ will be generally sufficient except where derivatives in $z$ will be invoked (in Section \ref{The viscous prescription}). 

  The angular momentum equation, averaged over turbulent fluctuations, reads \citep{BalbusPapaloizou1999}:
\begin{equation}
\rho \overline{u_R}\frac{\partial R\overline{u_\phi}}{\partial R}+\frac{1}{R}\frac{\partial}{\partial R}\left(R^2T_{R\phi}\right)+\frac{\partial}{\partial z}\left(RT_{z\phi}\right)=0,
\label{angular momentum equation}
\end{equation}
where $T_{R\phi}$ and $T_{z\phi}$, the $R\phi$ and $z\phi$ components of the turbulent stress tensor, are defined as:
\begin{eqnarray}
T_{i\phi}\equiv \rho\overline{\delta u_\phi\delta u_i}-\frac{\overline{B_\phi B_i}}{\mu_0},
\end{eqnarray}
with $\mathbf{B}$ the magnetic field and $i=R,z$. We parameterize $T_{R\phi}$ as:
\begin{equation}
T_{R\phi}(R,z)\equiv\frac{3}{2}\alpha (R,z) P(R,z).
\label{alpha}
\end{equation}
Note that equation (\ref{alpha}) is only a definition and does not presuppose any prescription of the stress tensor
. In particular, $\alpha$ may \textit{a priori} vary with $R$ and $z$. The arbitrary factor 3/2 has been introduced to facilitate direct comparison with the turbulent viscosity formalism.

  Equation (\ref{angular momentum equation}) may be rewritten as:
\begin{equation}
\overline{u_R}=-\frac{2}{\rho}\left[\frac{1}{R^{1/2}}\frac{\partial}{\partial R}\left(\frac{R^{1/2}}{\Omega}T_{R\phi}\right)+\frac{1}{\Omega}\frac{\partial T_{z\phi}}{\partial z}\right]
\label{uR}
\end{equation}
From now on, we will drop the overbars. The latter equation may be integrated to yield the vertical density-weighted average $\langle u_R \rangle_\rho$ 
 \citep[e.g.][]{LyndenBellPringle1974}\footnote{This assumes that $T_{z\phi}$ vanishes at infinity, which may not hold in case of steady outflows, but even then, its contribution to $\langle u_R\rangle_\rho$ is likely negligible \citep{BaiStone2012,Fromangetal2012}.}:
\begin{eqnarray}
\langle u_R\rangle_\rho \equiv \frac{1}{\Sigma}\int_{-\infty}^{+\infty}\rho u_R\mathrm{d}z & = & -\frac{2}{\Sigma R^{1/2}}\frac{\partial}{\partial R}\left(\frac{R^{1/2}}{\Omega}\int_{-\infty}^{+\infty}T_{R\phi}\mathrm{d}z\right)\nonumber\\
 & = &-\frac{3}{\Sigma R^{1/2}}\frac{\partial}{\partial R}\left(R^{1/2}\Sigma\langle\alpha\rangle_P\frac{c_s^2}{\Omega}\right).
\label{net velocity}
\end{eqnarray}
Thus, the net radial transport of gas depends only on $\langle\alpha\rangle_P$ as far as turbulence properties are concerned.

  We now turn to the two prescriptions for the Maxwell-Reynolds tensor to be considered in this paper, namely the ``viscous'' and the ``MHD'' prescription.

\subsection{The viscous prescription}
\label{The viscous prescription}
The viscous prescription, first introduced for hydrodynamical turbulence by  \citet{Boussinesq1877} and \citet{Reynolds1895} --- but see e.g. \citet{Schmitt2007} ---, and commonly used in protoplanetary disk contexts \citep[e.g.][]{Urpin1984,TakeuchiLin2002,Ciesla2009}, assumes that turbulence can be modeled as an effective kinematic viscosity $\nu$, so that: 
\begin{equation}
T_{R\phi}(R,z)=-\rho\nu R\frac{\partial}{\partial R}\left(\frac{u_\phi}{R}\right)
\label{TRphi viscous}
\end{equation}
\begin{equation}
T_{z\phi}(R,z)=-\rho\nu \frac{\partial u_\phi}{\partial z}= -\rho\nu\frac{\Omega}{2}\frac{\partial\mathrm{ln}T}{\partial R}z.
\end{equation}
where we have made use of equation (\ref{subkeplerian}).

  Comparing equation (\ref{TRphi viscous}) with equation (\ref{alpha}) leads to the identification:
\begin{equation}
\nu (R,z)=\alpha (R,z)\frac{c_s^2}{\Omega}.
\end{equation}
 At this point the viscous prescription is non-tautological only in the prescription of $T_{z\phi}$. It is generally further assumed that $\alpha$ is constant, at least vertically, i.e.: 
\begin{equation}
\alpha(R,z)=\alpha(R,0)=\langle \alpha\rangle_P.
\end{equation}
 The radial velocity may then be calculated from equation (\ref{uR}):
\begin{equation}
u_R(R,z)=-\frac{\alpha(R,0)c_s^2}{\Omega}\bigg[\frac{\partial}{\partial R}\mathrm{ln}\left(\left(\alpha(R,0)\Sigma\right)^3H\right)+\frac{\partial}{\partial R}\mathrm{ln}\left(\frac{c_s^5}{\Omega^3}\right)\left(\frac{z}{H}\right)^2\bigg].
\end{equation}
This flow, parabolic in $z$, is known as the meridional circulation (see Fig. \ref{velocity profile}). Typically, the velocity near the midplane is positive \citep{TakeuchiLin2002}, but the inward flows away from this region result in a positive net (vertically integrated) mass accretion rate (i.e. toward the Sun; see equation (\ref{net velocity})). This circulation may be interpreted as follows: at the midplane, the large radial density gradient ($\rho\propto\Sigma\Omega/c_s$) makes the viscous torque exerted from the inner, more rapidly rotating disk regions greater than that received from the outer, slower rotating disk, hence a net gain of angular momentum and an outward flow; at high altitudes, however, the density gradient is reduced and even switches sign, changing the direction of the effect \citep[e.g.][]{TakeuchiLin2002}.

\subsection{MRI-turbulent disk prescription}
\label{MHD prescription}

\begin{figure}
\resizebox{\hsize}{!}{
\includegraphics{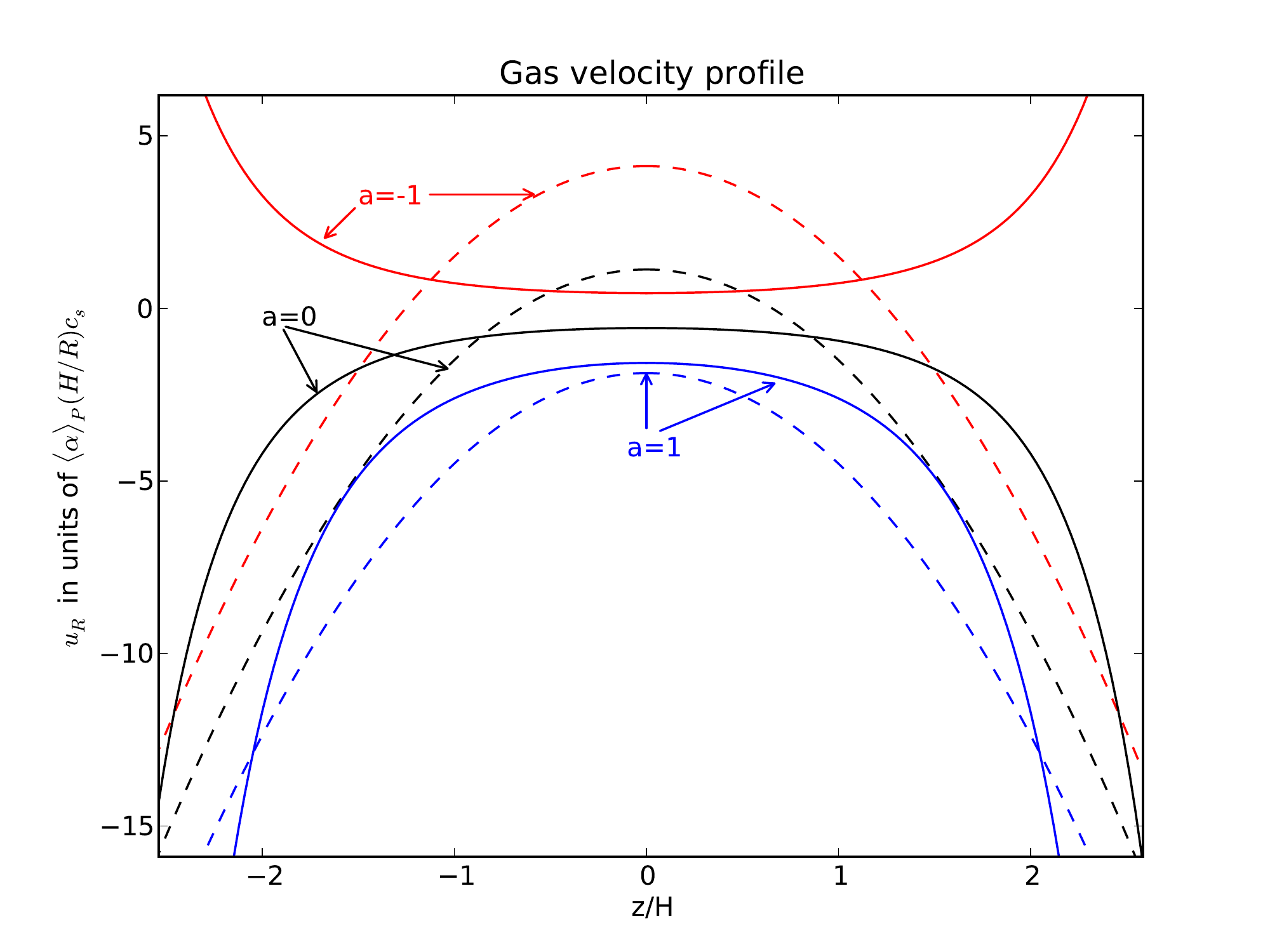}
}
\caption{Plot of the mean radial velocity of the gas for the viscous (dashed) and the MHD (continuous line) prescriptions. We have taken $\Sigma, T \propto R^{-0.75}$, and $\alpha(R,0)\propto R^a$ (with the local value of $\langle\alpha\rangle_P$ being $10^{-2}$), with three values of the exponent $a$ used in the plots. The viscous prescription entails a parabolic flow profile (the so-called meridional circulation) while the MHD prescription entails an inverted gaussian profile (until the corona, here at $|z|>z_{\rm max}=3.0H$). Here, for $a=-1$, the vertically integrated flow is outward in both prescriptions.}
\label{velocity profile}
\end{figure}

Simulations of magnetohydrodynamic turbulence have so far failed to exhibit a meridional circulation \citep{Fromangetal2011,Flocketal2011} and this may be traced to the discrepancy between the measured stress tensor and that assumed in the viscous prescription \citep{Fromangetal2011}. As noted by numerous numerical studies (\citet{Millerstone2000,Hiroseetal2006,Flaigetal2010,FromangNelson2006,Dzyurkevichetal2010,Sorathiaetal2010,Fromangetal2011,Flocketal2011,GuanGammie2011,Fromangetal2012,BaiStone2012}; see also \citet{Uzdensky2012}), the vertical profile of $T_{R\phi}$ shows a plateau for $|z|$ lower than a given $z_{\rm max}$. It then falls off, proportional to the density, in the corona, where magnetic pressure 
 is comparable to the gas pressure, quenching the MRI and leaving a transsonic turbulence there \citep[e.g.][]{Flocketal2011}. We thus adopt:
\begin{equation}
T_{R\phi}(R,z)=T_{R\phi}(R,0)\mathrm{exp}\left(\frac{z_{\rm max}^2-z^2}{2H^2}\theta(|z|-z_{\rm max})\right),
\label{TRphi MHD}
\end{equation}
with $\theta$ the Heaviside function, defined as:
\begin{equation}
\theta(x)=
\left\{\begin{array}{rr}
0\:\mathrm{if}\:x<0\\
1\:\mathrm{if}\:x\geq 0
\end{array}\right.
\end{equation}
We thus have 
\begin{equation}
\alpha(R,z)=
\left\{\begin{array}{rr}
\alpha(R,0)e^{z^2/2H^2}\:\mathrm{if}\:|z|\leq z_{\rm max}\\
\alpha_{\rm max}\:\mathrm{if}\:|z|\geq z_{\rm max}
\end{array}\right.
\label{alpha MRI}
\end{equation}
with
\begin{equation}
z_{\rm max}=H\sqrt{2\mathrm{ln}\left(\frac{\alpha_{\rm max}}{\alpha(R,0)}\right)}.
\end{equation}
As to $\alpha_{\rm max}$, we will adopt in numerical applications the fiducial value 1/3 that arises in the ideal MHD limit of the study of \citet{BaiStone2011}\footnote{We have recast their equations 23 and 25 in terms of our definition of $\alpha$ (equation (\ref{alpha})); $\alpha_{\rm max}$ corresponds here to the limit where gas and magnetic pressure are equal.}. This clearly differs from the viscous prescription.

We hence have:
\begin{equation}
\langle\alpha\rangle_P=\frac{2}{\sqrt{\pi}}\alpha(R,0)\sqrt{\mathrm{ln}\left(\frac{\alpha_{\rm max}}{\alpha(R,0)}\right)}+\alpha_{\rm max}\mathrm{erfc}\left(\sqrt{\mathrm{ln}\left(\frac{\alpha_{\rm max}}{\alpha(R,0)}\right)}\right).
\end{equation}
with erfc the complementary Gauss error function defined as:
\begin{equation}
\mathrm{erfc}(x)\equiv 1-\mathrm{erf}(x)\equiv \frac{2}{\sqrt{\pi}}\int_x^{\infty}e^{-y^2}\mathrm{d}y
\end{equation}
Note that we have made no assumption on the value of $\alpha(R,0)$ or on how it scales with other disk parameters as we focus here on the $z$ dependence of the flow. 

Considerably less is known about $T_{z\phi}$. We propose a break at $|z|=z_{\rm max}$ similar to that of $T_{R\phi}$ in the form:
\begin{equation}
T_{z\phi}(R,z)=C(R,z)\mathrm{exp}\left(\frac{z_{\rm max}^2-z^2}{2H^2}\theta(|z|-z_{\rm max})\right),
\label{Tzphi smooth}
\end{equation}
with $C(R,z)$ an as yet unspecified smooth function which symmetry about the midplane requires to be odd in $z$. 
By \textit{requiring} that $u_R$ as expressed by equation (\ref{uR}) be continuous at $z=\pm z_{\rm max}$, we have:
\begin{equation}
C(R,z_{\rm max})=-C(R,-z_{\rm max})=T_{R\phi}(R,0)\frac{\partial z_{\rm max}}{\partial R},
\end{equation}
which essentially sets the scale of $T_{z\phi}\sim (H/R)T_{R\phi}$.\footnote{Note that while our derivation makes $T_{z\phi}$ depend on the formally nonlocal quantity $\partial z_{\rm max}/\partial R$, this is not inconsistent with the fact that local simulations can exhibit a nonzero $T_{z\phi}$ (see e.g. \citealt{Fromangetal2012}) to which the radial gradient of $\alpha$ might adjust.} If we adopt the simplest dependence $C(R,z)\propto z$, we obtain:

\begin{equation}
T_{z\phi}(R,z)=T_{R\phi}(R,z)\frac{\partial\mathrm{ln}z_{\rm max}}{\partial R}z
\label{Tzphi linear}
\end{equation}
This yields a profile similar to that measured by \citet{Fromangetal2011}. 

  Equations (\ref{TRphi MHD}) and (\ref{Tzphi linear}) constitute what we will refer to as the ``MHD prescription''. It is noteworthy that it involves the same free parameters ($\alpha(R,0)$, or equivalently $\langle\alpha\rangle_P$, and its possible radial derivative) as the viscous one. 

  The mean radial velocity resulting from this prescription may then be calculated. For $|z|\leq z_{\rm max}$,
\begin{equation}
u_R(R,z)=-\frac{2}{\rho R^{1/2}z_{\rm max}}\frac{\partial}{\partial R}\left(R^{1/2}\frac{T_{R\phi}(R,0)}{\Omega}z_{\rm max}\right)=u_R(0)e^{\frac{z^2}{2H^2}},
\label{uR below corona}
\end{equation}
and for $|z|\geq z_{\rm max}$,
\begin{eqnarray}
u_R(R,z)=\frac{3\alpha_{\rm max}c_s^2}{\Omega}\Bigg[-\frac{\partial}{\partial R}\mathrm{ln}\left(\frac{R^{1/2}}{\Omega}\alpha_{\rm max}P(R,0)z_{\rm max}\right)\nonumber\\
+\frac{\partial}{\partial R}\mathrm{ln}\left(\frac{z_{\rm max}}{H}\right)\left(\frac{z}{H}\right)^2\Bigg].
\label{uR in corona}
\end{eqnarray}
The flow is plotted in Fig. \ref{velocity profile}. In contrast to the meridional circulation, velocities do not change sign (except perhaps in the corona if $\partial\alpha(R,0)/\partial R\neq 0$). Indeed the stress tensor no longer scales with the density vertically, in contrast to the viscous prescription, and thus the effect of the change in radial density gradient with height (see Section \ref{The viscous prescription}) disappears. Then, the radial velocity does not switch sign, and always has that of its average (at least outside the corona). The profiles are similar to those arising from the global simulations of \citet{Fromangetal2011}. They differ, however, from those of \citet{Flocketal2011}, which resemble an inverted meridional circulation profile, where the positive velocities in the corona correspond to outflows potentially escaping the disk. Such outflows launched from the disk 
 have been also recently studied in local simulations with uniform initial vertical magnetic field \citep{Lesuretal2012,BaiStone2012,Fromangetal2012} and if present, are thus not captured by our prescription of $T_{z\phi}$ (which would not vanish at infinity for outflow-launching simulations). We shall however argue later that its exact form has little effect on our conclusions. Our prescription for the Maxwell-Reynolds tensor is thus only intended here to provide a sensible MHD counterpart to the viscous one for comparison purposes. Certainly, more measurements of $T_{R\phi}$ \textit{and} $T_{z\phi}$ in stratified numerical simulations are needed to arrive at a more definitive prescription.


\section{Impact on the dynamics of solids}
\label{Impact on the dynamics of solids}

With the two above prescriptions on the gas flow in hand, we now investigate their impact on the dynamics of solids tightly coupled to the gas. 
We first briefly review the basic concepts of aerodynamic transport of solids in disks, before focusing on the vertical distribution of particles and their net radial motions. 
\subsection{Generalities}
Let us consider a population of identical spherical particles of internal density $\rho_s$ and radius $a$ embedded in the gas. We denote by $\rho_p$ the density of this population (viewed as a fluid) and
\begin{equation}
\Sigma_p=\int_{-\infty}^{+\infty}\rho_p\mathrm{d}z
\label{Sigma_p defined}
\end{equation}
their vertically integrated surface density. We also define a \textit{normalized solid-to-gas ratio} $f$ by:
\begin{equation}
\frac{\rho_p(R,z)}{\rho(R,z)}\equiv\frac{\Sigma_p(R)}{\Sigma(R)}f(R,z).
\label{f defined}
\end{equation}

The effects of gas drag is characterized by a stopping time $\tau$ \citep[see e.g.][]{Weidenschilling1977} from which one can define a measure of the coupling of the particles to the gas on an orbital timescale as
\begin{equation}
\textrm{St}\equiv\Omega\tau,
\end{equation}
which we will take to be $\ll 1$ in accordance with our tight coupling assumption.

Neglecting any feedback of the solids on the gas, the continuity equation averaged over turbulent fluctuations reads:
\begin{eqnarray}
\frac{\partial\rho_p}{\partial t}+\frac{1}{R}\frac{\partial}{\partial R}\left[R\left(\rho_pv_{\rm p, R}-D_{RR}\rho\frac{\partial}{\partial R}\left(\frac{\rho_p}{\rho}\right)\right)\right]\nonumber\\+\frac{\partial}{\partial z}\left[\rho_pv_{\rm p, z}-D_{zz}\rho\frac{\partial}{\partial z}\left(\frac{\rho_p}{\rho}\right)\right]=0,
\label{continuity 2D}
\end{eqnarray}
where we have introduced the mean velocity $\mathbf{v}_p$ of the particles given by \citep{YoudinGoodman2005}
\begin{equation}
\mathbf{v}_p\equiv\mathbf{u}+\mathbf{v}_{\rm drift}=\mathbf{u}+\tau\frac{\nabla P}{\rho},
\label{TVA}
\end{equation}
and also diffusion coefficients $D_{RR}$ and $D_{zz}$ parameterized as:
\begin{eqnarray}
D_{RR}=\delta_R\frac{c_s^2}{\Omega}\:\:\:\:\:\:\mathrm{and}\:\:\:\:\:\:
D_{zz}=\delta_z\frac{c_s^2}{\Omega},
\end{eqnarray}
with $\delta_R$ and $\delta_z$ dimensionless parameters of order $\alpha$
. We will assume that the radial and vertical Schmidt numbers
\begin{eqnarray}
\textrm{Sc}_R\equiv\frac{\alpha}{\delta_R}\:\:\:\:\:\:\mathrm{and}\:\:\:\:\:\:
\textrm{Sc}_z\equiv\frac{\alpha}{\delta_z}
\end{eqnarray}
are vertically constant.


  Following \citet{Jacquetetal2012S}, we coin:
\begin{equation}
S\equiv\frac{\textrm{St}}{\alpha}\sim \frac{v_{\rm drift,R}}{u_R}\:\:\:\:\:\:\mathrm{and}\:\:\:\:\:\:S_z\equiv\frac{\textrm{St}}{\delta_z}=S\times \textrm{Sc}_z,
\label{S}
\end{equation}
which measure the coupling between particles and gas on a global scale, in the radial and vertical directions, respectively.

\subsection{Vertical distribution}

\begin{figure}
\resizebox{\hsize}{!}{
\includegraphics{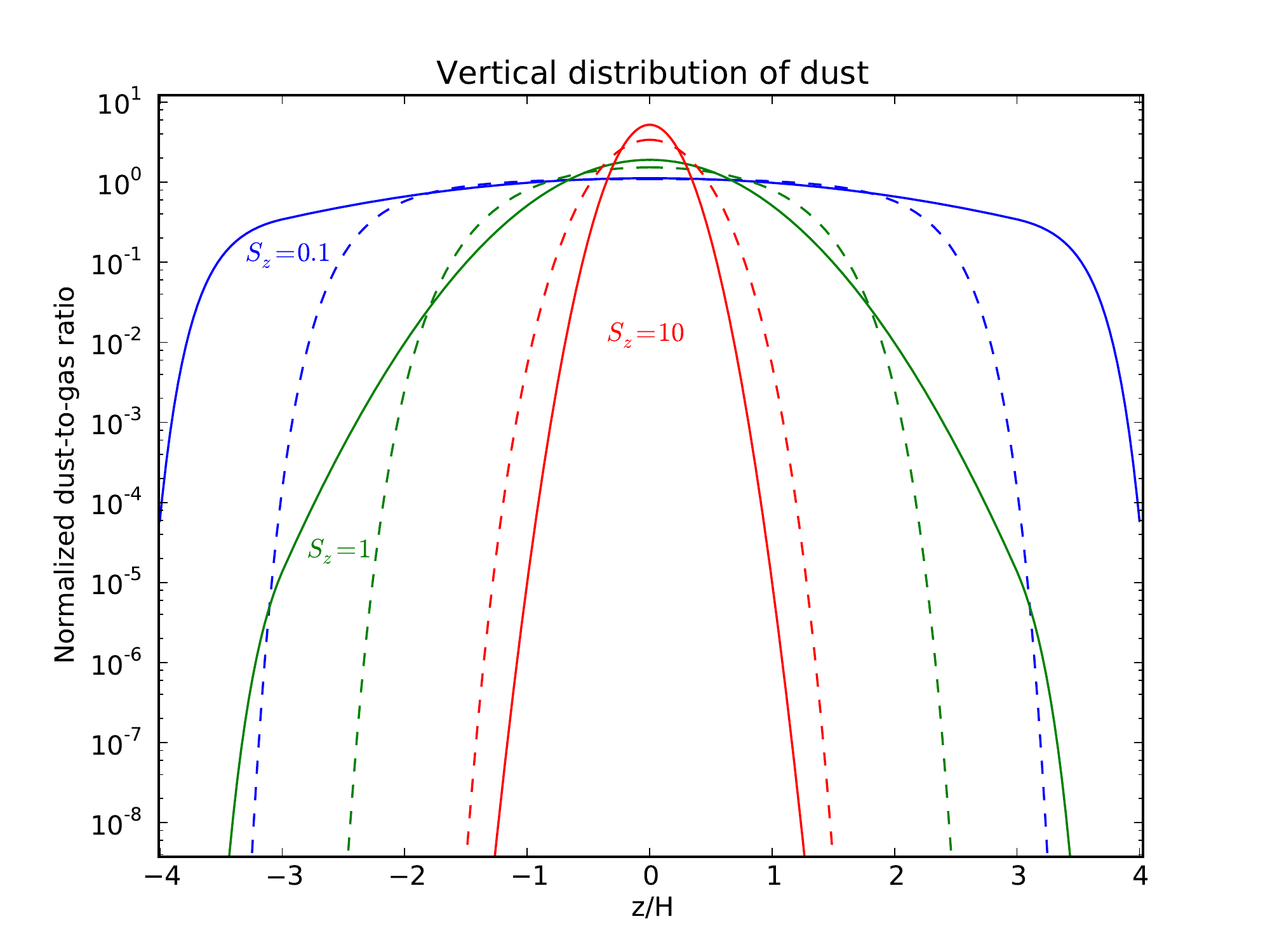}
}
\caption{Plot of the dust-to-gas ratio normalized to the column density ratio ($f$) as a function of $z$, for both viscous (dashed line) and MRI-turbulent prescriptions (solid line) and for three values of the ``settling parameter'' $S_z\equiv \textrm{St}/\langle\delta_z\rangle_P$, with $\langle\delta_z\rangle_P=10^{-2}$. In the viscous prescription, this ratio falls off as a ``double gaussian'' whereas it only does so as a gaussian for the MHD prescription over the bulk of the disk thickness, before decreasing as a ``double gaussian'' in the corona (here at $|z|>z_{\rm max}=3.0H$).}
\label{settling}
\end{figure}

Equation (\ref{continuity 2D}) is dominated by the vertical flux balance between settling and diffusion. On timescales longer than the vertical mixing timescale 
\begin{equation}
t_{\rm vm}=\frac{1}{\Omega\:\mathrm{max}(\delta_z,\textrm{St})},
\end{equation}
the vertical distribution of the particles obeys:
\begin{equation}
\frac{\partial\mathrm{ln}f}{\partial z}=-S_z(R,z)\frac{z}{H^2},
\label{vertical equilibrium}
\end{equation}
where the role of $S_z$ as a measure of settling is apparent. It may be noted that our assumption of constant Sc$_z$ makes the vertical distribution independent of the prescription of $T_{z\phi}$.

  In the viscous prescription, where $\alpha$, and hence $\delta_z$ is vertically constant, $S_z\propto \rho^{-1}$ so that equation (\ref{vertical equilibrium}) may be integrated as \citep{TakeuchiLin2002}:
\begin{eqnarray}
f(R,z)&=&\frac{\sqrt{2\pi}H}{\int_{-\infty}^{+\infty}\exp{(-S_z(R,z')-z'^2/2H^2)}\mathrm{d}z'}\exp{(-S_z(R,z))}\nonumber\\
&\propto&\exp{(-S_z(R,0)e^{\frac{z^2}{2H^2}})}.
\label{f constant deltaz}
\end{eqnarray}
a ``double gaussian'' which drops rapidly at large $|z|$.

  In the MHD prescription, $S_z$ is vertically constant for $|z|\leq z_{\rm max}$ before increasing as $\rho^{-1}$ in the corona. This gives:
\begin{equation}
f(R,z)=
f(R,0)\mathrm{exp}\left(-S_z(R,0)\frac{z^2}{2H^2}\right)
\end{equation}
for $|z|\leq z_{\rm max}$, and
\begin{equation}
f(R,0)\left(\frac{\alpha(R,0)}{\alpha_{\rm max}}\right)^{S_z(R,0)}\mathrm{exp}\left(S_z(R,0)\left(1-\frac{\alpha(R,0)}{\alpha_{\rm max}}e^{z^2/2H^2}\right)\right)
\end{equation}
for $|z|> z_{\rm max}$, with
\begin{eqnarray}
f(R,0)&=  \Bigg[\frac{\mathrm{erf}\left(\sqrt{(S_z(R,0)+1)\mathrm{ln}(\alpha_{\rm max}/\alpha(R,0))}\right)}{\sqrt{S_z(R,0)+1}}
+\frac{2}{\sqrt{\pi}}\left(\frac{\alpha(R,0)}{\alpha_{\rm max}}\right)^{S_z(R,0)}\nonumber\\ & \int_{\sqrt{\mathrm{ln}(\alpha_{\rm max}/\alpha(R,0))}}^{+\infty}\mathrm{exp}\left(-y^2+S_z(R,0)\left(1-\frac{\alpha(R,0)}{\alpha_{\rm max}}e^{y^2}\right)\right)\mathrm{d}y\Bigg]^{-1}
\end{eqnarray}
At first the solid-to-gas ratio falls off only as a gaussian (because of increasing $\alpha$ counteracting looser coupling due to decreasing $\rho$) and only in the corona does it decrease as a ``double gaussian'' similarly to the viscous prescription case. Thus, a purely gaussian fit tends to overestimate, and a ``double gaussian'' one to underestimate the dust density away from the midplane in this case, as observed by \citet{FromangNelson2009} in their simulations of dust settling in MRI-driven turbulence.

The vertical distribution in both prescriptions is plotted in Fig. \ref{settling}. Similarly to \citet{FromangNelson2009}, we suggest that observations of present-day protoplanetary disks, in probing dust remaining on their surface, could discriminate between the two vertical distributions.

\subsection{Net radial flow}

\begin{figure}
\resizebox{\hsize}{!}{
\includegraphics{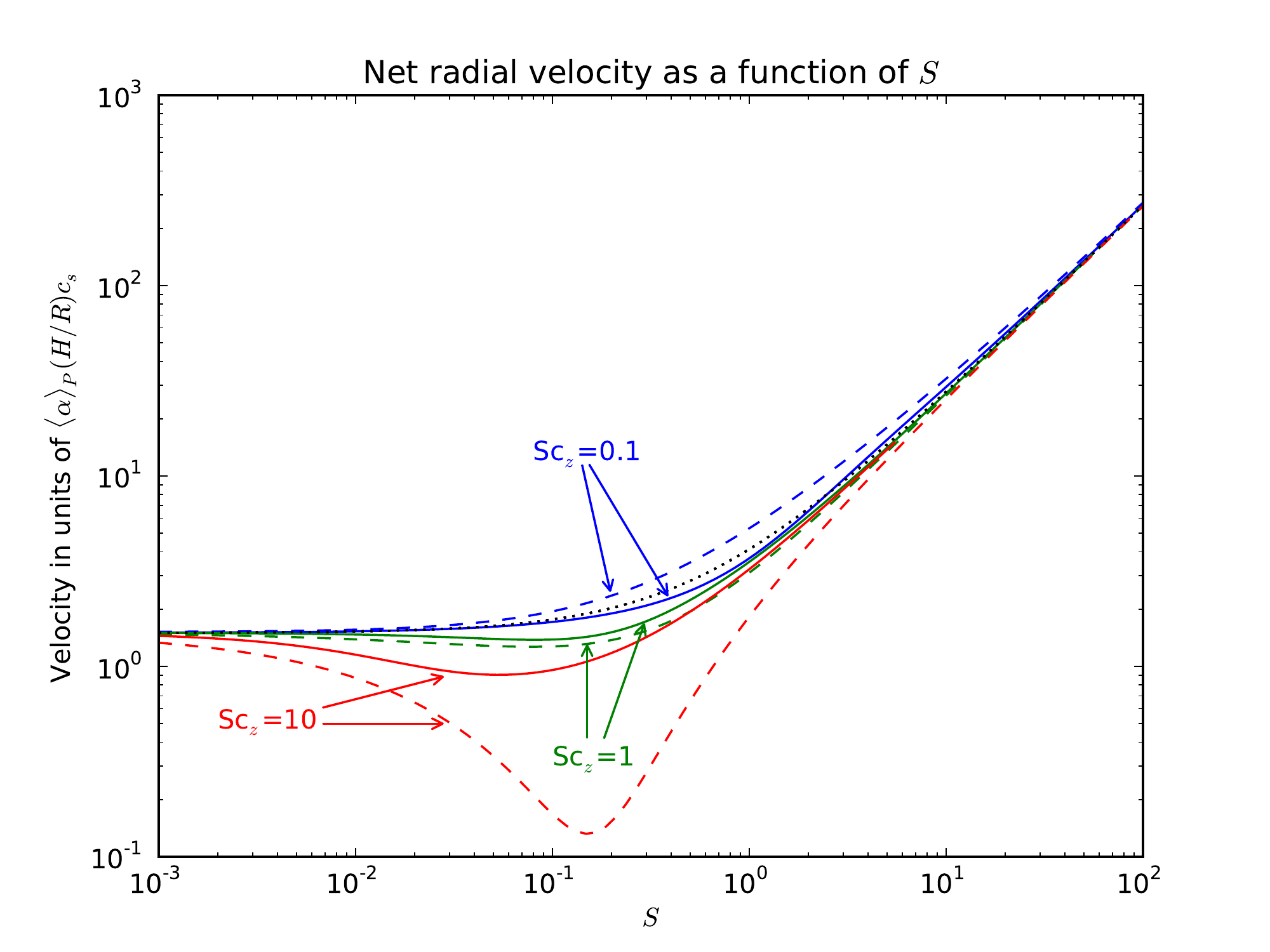}
}
\caption{Plot of the net vertically averaged velocity $v_{\rm p, 1D}$ 
 as a function of $S=\textrm{St}(R,0)/\langle\alpha\rangle_P$ for the viscous prescription (dashed line) and the MHD prescription (solid line), for three values of Sc$_z$ (0.1 in blue, 1 in green and 10 in red). The velocities are always inward and we plot here their absolute value. We have taken $\Sigma, T\propto R^{-0.75}$, $\partial\alpha(R,0)/\partial R=0$, $\delta_R=\delta_z$ and $\langle\alpha\rangle_P=10^{-2}$. The dotted line corresponds to the simple one-dimensional approximation (see equation (\ref{1D approximation})) which is common to both prescriptions and independent of Sc$_z$. The approximation is arbitrarily good in the limits $S\ll 1$ and $S\gg 1$ and the imparted error does not exceed a factor of two when $S\sim 1$ except for the viscous prescription at Sc$_z$=10 where it can reach one order of magnitude.}
\label{velvert}
\end{figure}

We now turn to the radial motions of the particles. Specifically, we want to integrate the continuity equation (\ref{continuity 2D}) over $z$ and deduce the net radial flow for both prescriptions. Formally, the result of this integration reads:
\begin{equation}
  \frac{\partial\Sigma_p}{\partial t}+\frac{1}{R}\frac{\partial}{\partial R}\left[R\left(v_{\rm p,1D}\Sigma_p-\langle D_{RR}\rangle_{\rho_p}\Sigma\frac{\partial}{\partial R}\left(\frac{\Sigma_p}{\Sigma}\right)\right)\right]=0,
\label{continuity 1D}
\end{equation}
with
\begin{equation}
v_{\rm p,1D}=\langle u_R\rangle_{\rho_p}+\langle v_{\rm drift}\rangle_{\rho_p}+v_{\rm p, corr},
\label{vp1D}
\end{equation}
where
\begin{equation}
v_{\rm p, corr}=-\int_{-\infty}^{+\infty}D_{RR}\frac{\partial f}{\partial R}\frac{\rho\mathrm{d}z}{\Sigma}
\label{vpcorr}
\end{equation}
is a correction from the diffusion term, which is generally negative (see e.g. equation (\ref{diffusive correction})), as the disk thickens with increasing heliocentric distance, inducing its ``surface'' to diffuse particles inward into the rarefied, particle-depleted gas. 

  If we restrict interest to timescales longer than $t_{\rm vm}$, we can adopt the equilibrium vertical distributions of the previous subsection to calculate the above. Results are presented in appendix \ref{appendix exact} and $v_{\rm p, 1D}$ is plotted in Fig. \ref{velvert} for different values of the vertical Schmidt number. The main conclusion to draw from this plot is this net velocity is weakly dependent on the exact flow structure and is generally well approximated by
\begin{equation}
v_{\rm p, 1D}(R)\approx\langle u_R \rangle_\rho + v_{\rm drift, R}(R,0),
\label{1D approximation}
\end{equation}
a widespread formula in one-dimensional calculations \citep[e.g.][]{Cuzzietal2003,Ciesla2010,HughesArmitage2010,YangCiesla2012}. Agreement is arbitrarily good when $S$ is either very large or very small, and the deviation is largest for $S\sim 1$. The deviation is generally less than a factor of 2, except for the viscous prescription \textit{if} the Schmidt number is large.  But even then, the net radial velocity does not (in general) become positive, as had been noted by \citet{TakeuchiLin2002}, even if they did not include the diffusion correction. 

  Why does the 1D approximation (\ref{1D approximation}) provide a so good match, at least when $\textrm{Sc}_R\sim\textrm{Sc}_z\sim 1$? This stems from the facts that (see also \citealt{Jacquetetal2012S}): 
\begin{itemize}
\item[(i)] In the limit $S, S_z\ll 1$, particles are well-mixed with the gas and only drift negligibly with respect to it, so that their vertically-averaged velocity is indistinguishable from that of the gas, $\langle u_R \rangle_\rho$.
\item[(ii)] In the limit $S, S_z\gg 1$, particles are concentrated at the midplane and their motion is dominated by drag-related drift (evaluated in this region), so that their net radial velocity is essentially $v_{\rm drift}(R,0)$.
\end{itemize}
As the expression (\ref{1D approximation}) captures both limiting behaviors---with diffusion correction $v_{\rm p, corr}$ being negligible in both limits \citep{Jacquetetal2012S}---, it indeed provides a suitable approximation for all possible values of $S$. It is noteworthy that the above reasoning is general and extends to prescriptions other than the ones considered in this study, e.g. if one adopts an alternative mathematical expression for $T_{z\phi}$ in the MHD prescription.\footnote{The contribution to $\langle u_R\rangle_{\rho_p}$ of $T_{z\phi}$ in our prescription is:
\begin{eqnarray}
-3\frac{c_s^2}{\Omega}\frac{\partial\mathrm{ln}z_{\rm max}}{\partial R}\left(\langle\alpha\rangle_{\rho_p}-\sqrt{\frac{2}{\pi}}\alpha_{\rm max}\int_{z_{\rm max}}^{+\infty}\left(\frac{z}{H}\right)^2f(R,z)e^{-z^2/2H^2}\frac{\mathrm{d}z}{H}\right)\nonumber
\end{eqnarray}
zero for $S \ll 1$ and $-3(\alpha(R,0) c_s^2/\Omega)\partial\mathrm{ln}z_{\rm max}/\partial R \ll \langle v_{\rm drift} \rangle_{\rho_p}$ for $S\gg 1$.
}

  In detail, some deviation is expected (and seen in Fig. \ref{velvert}) for $S\sim 1$ which is outside either above limit, but as the approximations just begin to break down there, the order of magnitude at least should be accurate. If however Sc$_z$ is very different from unity, the domain of non-validity of the above regimes is wider, and depending on the vertical variations of the velocities, this could impart a more significant error. This is what we see for the viscous prescription if Sc$_z\gg 1$. Indeed, under such conditions, one can have $S<1<S_z$: then, the particles are concentrated around the midplane (since $S_z>1$) so that their gas velocity has an important contribution from the outward-directed flows, which is incompletely compensated by the inward drag-related drift (since $S\sim v_{\rm drift}/u_R<1$), hence a significantly less negative $v_{\rm p, 1D}$. We note however that large values of the Schmidt number are not expected in the hydrodynamical turbulence which the viscous prescription is intended to model \citep{Prinn1990,DubrulleFrisch1991}, in contrast to MHD turbulence \citep{Carballidoetal2005,JohansenKlahr2005,Johansenetal2006}, so that this situation appears unlikely anyway.

\section{Discussion}
\label{Discussion}

  From the preceding section, it appears that the net radial transport of particles is weakly sensitive to the vertical flow structure and is well approximated by the one-dimensional expression given by equation (\ref{1D approximation}). In particular, \textit{if} the mass accretion rate is \textit{positive} (toward the Sun), the non-diffusive contribution to the net radial flux should be generally inward.

  This might seem paradoxical in the case of the viscous prescription, as then, a population of small ($S\ll 1$) particles around the midplane \textit{should} be transported outward, even in the absence of turbulent diffusion. The paradox lies in our calculation pertaining to timescales longer than the vertical mixing timescale $t_{\rm vm}$ (to warrant use of the equilibrium vertical distribution of particles) and thus to radial scales larger than the corresponding radial excursion
\begin{equation}
\Delta R_{\rm vm}=\mathrm{max}\left(\sqrt{D_{RR}t_{\rm vm}},|v_{\rm p, R}|t_{\rm vm}\right)\sim \mathrm{max}\left(\frac{H}{\sqrt{1+S}},\frac{H^2}{R}\right)<H. 
\end{equation} 
So while a finger of material lying near the midplane would undergo some outward transport in the viscous prescription, it would not travel significantly further than this \textit{if we ignore radial diffusion}. This we can see perhaps more concretely with a simple toy model, which we now present (see also Fig. \ref{toy meridional}):

\begin{figure}
\resizebox{\hsize}{!}{
\includegraphics{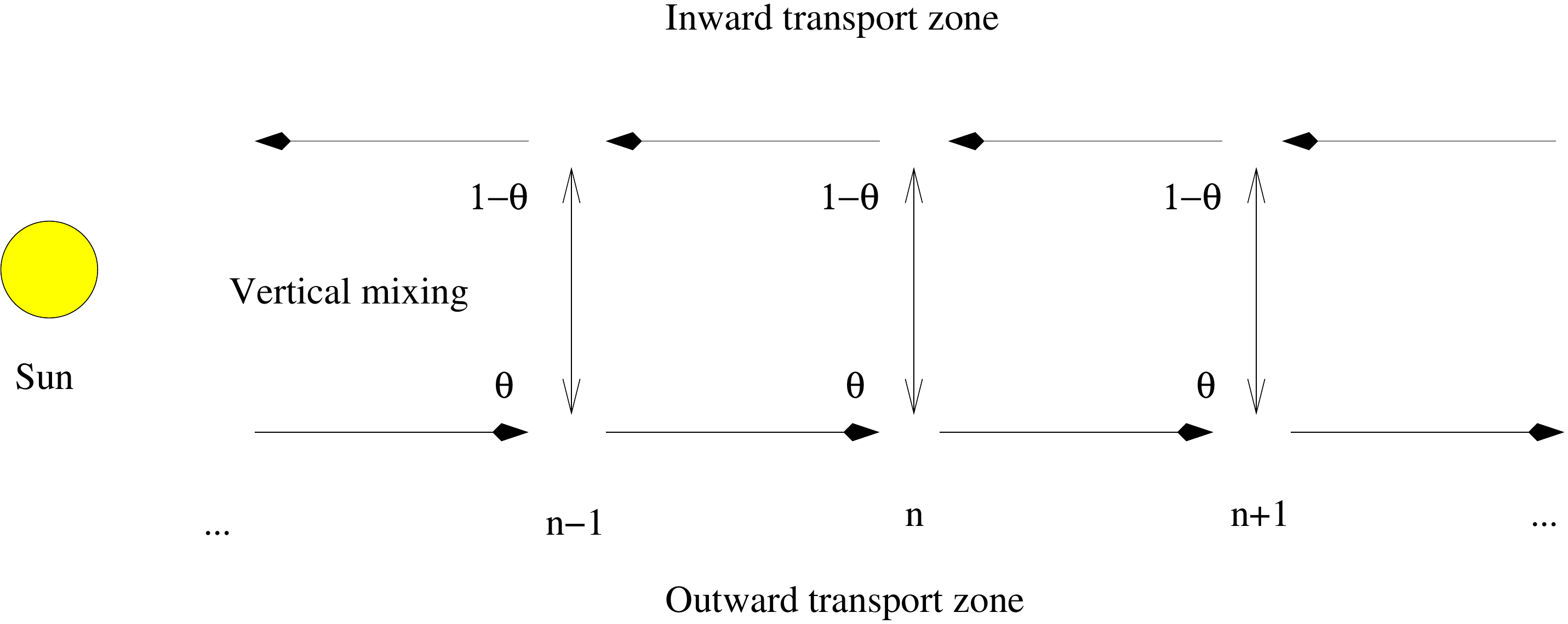}
}
\caption{Cartoon of the toy model for meridional circulation. We distinguish between an ``outward transport zone'' (rightward arrows) and an ``inward transport zone'' (leftward arrows), which are assigned fractions $\theta$ and $1-\theta$ ($\theta < 1/2$) of the local column density, respectively. Integers $n-1$, $n$, $n+1$ represent radial locations whose separation corresponds to one vertical mixing timescale. Vertical mixing is symbolized by double vertical arrows.}
\label{toy meridional}
\end{figure}

  We consider the disk thickness to consist in an ``outward transport zone'' (the midplane), comprising a fraction $\theta$ of material, and an ``inward transport zone'' (the ``atmosphere''). We ignore radial diffusion and consider that both zones transport material at equal and opposite velocities, so that the net flow is inward if $\theta < 1/2$, as we shall henceforth assume.  We 
discretize time and space (in the radial direction), with the temporal and radial steps being $t_{\rm vm}$ and $\Delta R_{\rm vm}$, respectively. 
Each time step corresponds to two successive motions: advection one radial step forward or backward within each zone, and vertical mixing between the two zones. Thus, if we denote by $u_{\rm n, t}$ the total column density of a contaminant at time $t$ and at radial location $n$---both natural integers, in units of $t_{\rm vm}$ and $\Delta R_{\rm vm}$---, $u_{\rm n,t+1}$ is given by: 
\begin{equation}
u_{\rm n,t+1}=\theta u_{\rm n-1,t}+(1-\theta)u_{\rm n+1,t}.
\label{recurrence}
\end{equation}
As boundary condition, we set $u_{0,t}=1$ (in some arbitrary unit) and as initial condition $u_{n,0}=0$ for $n\geq 1$.


  One can demonstrate\footnote{By recurrence on $t$ for (i)-(iii) and noting that (ii)-(iii) implies convergence of $u_{n,t}$ and that the limit has to be a stationary solution of the recurrence equation, but must have zero projection on the solution $(1)_{n\in\mathbb{N}}$.} that $u_{\rm n,t}$ determined by equation (\ref{recurrence}) and its boundary and initial conditions satisfies: 
\begin{itemize}
\item[(i)] $u_{\rm n,t}$ is a monotonically decreasing function of $n$. 
\item[(ii)] $u_{\rm n,t}$ is a monotonically increasing function of $t$. 
\item[(iii)] $0\leq u_{\rm n,t}\leq (\theta/(1-\theta))^n$ (with the upper bound being a stationary solution of equation (\ref{recurrence})).
\item[(iv)] $u_{\rm n,t}$ converges toward $(\theta/(1-\theta))^n$ as $t\rightarrow +\infty$. 
\end{itemize}
 
  Since $\theta<1/2$, $(\theta/(1-\theta))^n$ and \textit{a fortiori} $u_{\rm n,t}$ vanish as $n\rightarrow+\infty$ and in practice become negligibly small after several radial steps, corresponding, if we revert to physical units, to a radial excursion much smaller than $R$. 
The conclusion from this toy model is thus that no significant outward transport can be achieved without including radial diffusion, even in a meridional circulation context.

  However, 2D disk models assuming the viscous prescription did include radial diffusion, and did achieve significant outward transport \citep[e.g.][]{Ciesla2009}, but from the above, the latter should be viewed as being \textit{caused by the turbulent diffusion} rather than the meridional circulation itself. Nonetheless, it does hold that the meridional circulation, in reducing somewhat the inward advective velocities (see Fig. \ref{velvert}) which counteract turbulent diffusion, enhances outward transport relative to the one-dimensional approximation, and this can have nonnegligible effects for long-range transport. For example, in the ``outward transport'' simulation of \citet{Ciesla2009} with $\dot{M}=10^{-6}\:\rm M_\odot/yr$, a crystalline fraction 
 of 40 \% is obtained (after $10^5$ yr) at 10 AU, where  the one-dimensional approximation would have predicted 17 \% (using equation 29 of \citet{Jacquetetal2012S}). 
However, the one-dimensional model \textit{would} have retrieved a crystallinity of 40 \% had we taken Sc$_R=0.5$ instead of the Sc$_R=1$ chosen by \citet{Ciesla2009}, which is at least equally plausible. Thus, the differences between meridional circulation and the one-dimensional approximation circulation are actually within the errors of turbulence parameters
. A corollary is that the occurrence of high-temperature materials in the samples returned from comet Wild 2 \citep{Zolenskyetal2006} is \textit{not} diagnostic \textit{per se} of meridional circulation. More generally, \textit{net radial transport of early solar system material is no compelling constraint on the vertical variations of the gas flow}. The question of the form of the Maxwell-Reynolds tensor in protoplanetary disks, be it in MRI-active regions or in the dead zone, is thus most appropriately addressed by numerical simulations of gas turbulence, although, as we mentioned earlier, observations of protoplanetary disks could also offer important diagnostics linked to settling.
 
  We caution before closing this section that we have not considered, in this study, the possibility of outflows launched from the disk \citep[e.g.][]{Flocketal2011,BaiStone2012,Fromangetal2012,Lesuretal2012}, which would yield a sink term on the right-hand-side of equation (\ref{continuity 1D})---potentially important for vertically well-mixed particles ($S_z\ll 1$). The possibility and quantitative importance of such outflows has yet to be explored in more details before their impact on the radial motion of solids can be satisfactorily addressed.

\section{Conclusion}
\label{Conclusion}

In this paper, we have considered two possible flow structures in the protoplanetary disk, which correspond to two different prescriptions of the turbulent Maxwell-Reynolds tensor: 
\begin{itemize}
\item[(i)] A \textit{viscous prescription}, commonly used in the litterature, which typically gives rise to the so-called meridional circulation with outward flows around the midplane and inward flows away from it.
\item[(ii)] A \textit{MHD prescription}, which we introduce and motivate, appropriate for magnetohydrodynamical turbulence (without outflows) and where velocities retain the same sign over the bulk of the disk's thickness.
\end{itemize}

  We have compared their effects on the dynamics of small solids. While, in the vertical direction, the distribution falls off less steeply (as a gaussian) in the MHD prescription than in the viscous one, it is found that the net radial flux differs little between the two. In fact, this radial flux is well approximated by a commonly used one-dimensional formula, which only depends on the vertically averaged turbulence parameter $\langle\alpha\rangle_P$. We have shown that this can be generally expected for any likely prescription of the flow (and not only those considered here). 

  It thus follows that in itself, evidence of outward transport of inner solar system material to the comet-forming region cannot be viewed as diagnostic of meridional circulation, whose physical reality is still controversial. In fact, if the mass accretion rate is positive, this outward transport is rather to be attributed to turbulent diffusion. More generally, the net radial transport of early solar system material does not significantly constrain the flow structure of the gas over the disk's thickness.

\begin{appendix}

\section{Exact results for vertically-integrated radial particle flux in the viscous and MHD prescription}
\label{appendix exact}

We list here exact results for different quantities allowing one to calculate the net radial flux of solids in the two prescriptions, assuming an equilibrium vertical distribution of the said solids.


\subsection{Viscous prescription}

\begin{equation}
\langle D_{RR}\rangle_{\rho_p}=D_{RR}(R,0)
\end{equation}
\begin{equation}
\langle u_R\rangle_{\rho_p}=u_R(R,0)-\frac{\alpha(R,0)c_s^2}{\Omega}\frac{\partial}{\partial R}\mathrm{ln}\left(\frac{c_s^5}{\Omega^3}\right)\langle\left(\frac{z}{H}\right)^2\rangle_{\rho_p}
\end{equation}
\begin{equation}
\langle v_{\rm drift,R}\rangle_{\rho_P}=v_{\rm drift,R}(R,0)\frac{\int_0^{+\infty}\left(1+2\frac{\partial\mathrm{ln}H}{\partial\mathrm{ln}P}y^2\right)\mathrm{exp}\left(-S_z(R,0)e^{y^2}\right)\mathrm{d}y}{\int_0^{+\infty}\mathrm{exp}\left(-S_z(R,0)e^{y^2}-y^2\right)\mathrm{d}y}
\end{equation}
\begin{equation}
v_{\rm p,corr}=D_{RR}\frac{\partial\mathrm{ln}H}{\partial R}\left(\langle\left(\frac{z}{H}\right)^2\rangle_{\rho_p}-1\right).
\label{diffusive correction}
\end{equation}
with
\begin{equation}
\frac{\partial\mathrm{ln}H}{\partial\mathrm{ln}P}\equiv\frac{\partial\mathrm{ln}H/\partial R}{\partial\mathrm{ln}P(R,0)/\partial R}
\end{equation}

\begin{equation}
\langle\left(\frac{z}{H}\right)^2\rangle_{\rho_p}=2\frac{\int_0^{+\infty}y^2\mathrm{exp}\left(-S_{z}(R,0)e^{y^2}-y^2\right)\mathrm{d}y}{\int_0^{\infty}\mathrm{exp}\left(-S_z(R,0)e^{y^2}-y^2\right)\mathrm{d}y}
\end{equation}

\subsection{MHD prescription}

\begin{eqnarray}
\langle D_{RR}\rangle_{\rho_p}=D_{RR}(R,0)\bigg[f(R,0)\frac{\mathrm{erf}\left(\sqrt{S_z(R,0)\mathrm{ln}(\alpha_{\rm max}/\alpha(R,0))}\right)}{\sqrt{S_z(R,0)}}\nonumber\\
+\frac{\alpha_{\rm max}}{\alpha(R,0)}\left(1-f(R,0)\frac{\mathrm{erf}\left(\sqrt{(S_z(R,0)+1)\mathrm{ln}(\alpha_{\rm max}/\alpha(R,0))}\right)}{\sqrt{S_z(R,0)+1}}\right)\bigg]
\end{eqnarray}
\begin{eqnarray}
\langle u_R\rangle_{\rho_p}=u_R(R,0)f(R,0)\bigg[\frac{\mathrm{erf}\left(\sqrt{S_z(R,0)\mathrm{ln}(\alpha_{\rm max}/\alpha(R,0))}\right)}{\sqrt{S_z(R,0)}}\nonumber\\
+\frac{2A}{\sqrt{\pi}}\left(\frac{\alpha(R,0)}{\alpha_{\rm max}}\right)^{S_z(R,0)-1}
\bigg]\nonumber\\
-3\frac{\alpha_{\rm max}c_s^2}{\sqrt{\pi}\Omega}\frac{\partial\mathrm{ln}\alpha(R,0)}{\partial R}f(R,0)\Bigg[\frac{1}{\mathrm{ln}\left(\alpha_{\rm max}/\alpha(R,0)\right)}\nonumber\\
\bigg(B+\frac{\alpha(R,0)}{\alpha_{\rm max}}\left(\sqrt{\mathrm{ln}\left(\frac{\alpha_{\rm max}}{\alpha(R,0)}\right)}-2S_z(R,0)C\bigg)\right)\nonumber\\
+2A\left(\frac{\alpha(R,0)}{\alpha_{\rm max}}\right)^{S_z(R,0)}
\Bigg]
\end{eqnarray}

\begin{eqnarray}
\langle v_{\rm drift,R}\rangle_{\rho_p}=v_{\rm drift,R}(R,0)f(R,0)\Bigg[
\frac{\mathrm{erf}\left(\sqrt{S_z(R,0)\mathrm{ln}\left(\alpha_{\rm max}/\alpha (R,0)\right)}\right)}{\sqrt{S_z(R,0)}}\nonumber\\+\frac{2B}{\sqrt{\pi}}
+\frac{\partial\mathrm{ln}H}{\partial\mathrm{ln}P}\Bigg(
\frac{1}{S_z(R,0)}\bigg(\frac{\mathrm{erf}\left(\sqrt{S_z(R,0)\mathrm{ln}\left(\alpha_{\rm max}/\alpha(R,0)\right)}\right)}{\sqrt{S_z(R,0)}}\nonumber\\
-2\sqrt{\frac{\mathrm{ln}\left(\alpha_{\rm max}/\alpha(R,0)\right)}{\pi}}\left(\frac{\alpha(R,0)}{\alpha_{\rm max}}\right)^{S_z(R,0)}\bigg)
+\frac{4C}{\sqrt{\pi}}\left(\frac{\alpha(R,0)}{\alpha_{\rm max}}\right)^{S_z(R,0)}
\Bigg)
\Bigg]
\end{eqnarray}

\begin{eqnarray}
v_{\rm p, corr}=-\frac{\partial\mathrm{ln}f(R,0)}{\partial R}\langle D_{RR}\rangle_{\rho_p}\nonumber\\
-D_{RR}(0)S_z(R,0)f(R,0)\Bigg[\left(\frac{\partial\mathrm{ln}\Sigma}{\partial R}+\frac{\partial\mathrm{ln}\alpha(R,0)}{\partial R}\right)\nonumber\\\Bigg(
\frac{1}{S_z(R,0)}\Bigg(\frac{\mathrm{erf}\left(\sqrt{S_z(R,0)\mathrm{ln}\left(\alpha_{\rm max}/\alpha(R,0)\right)}\right)}{\sqrt{S_z(R,0)}}\nonumber\\
-2\sqrt{\frac{\mathrm{ln}\left(\alpha_{\rm max}/\alpha(R,0)\right)}{\pi}}\left(\frac{\alpha(R,0)}{\alpha_{\rm max}}\right)^{S_z(R,0)}\Bigg)\nonumber\\
+
\mathrm{ln}\left(\frac{\alpha_{\rm max}}{\alpha(R,0)}\right)
\left(\frac{\alpha(R,0)}{\alpha_{\rm max}}\right)^{S_z(R,0)-1}\frac{2A}{\sqrt{\pi}}
\Bigg)\nonumber\\
+\frac{\partial\mathrm{ln}\Sigma}{\partial R}\frac{2}{\sqrt{\pi}}\left(\frac{\alpha(R,0)}{\alpha_{\rm max}}\right)^{S_z(R,0)}\left(B-A\frac{\alpha_{\rm max}}{\alpha(R,0)}\right)\nonumber\\
+\frac{\partial\mathrm{ln}H}{\partial R}\Bigg(
\frac{1}{S_z(R,0)}\Bigg(\frac{\mathrm{erf}\left(\sqrt{S_z(R,0)\mathrm{ln}\left(\alpha_{\rm max}/\alpha(R,0)\right)}\right)}{\sqrt{S_z(R,0)}}\nonumber\\-2\sqrt{\frac{\mathrm{ln}\left(\alpha_{\rm max}/\alpha(R,0)\right)}{\pi}}\left(\frac{\alpha(R,0)}{\alpha_{\rm max}}\right)^{S_z(R,0)}\Bigg)
+\frac{4C}{\sqrt{\pi}}\left(\frac{\alpha(R,0)}{\alpha_{\rm max}}\right)^{S_z(R,0)}
\Bigg)\Bigg]
\nonumber\\
\end{eqnarray}

with
\begin{equation}
A=\int_{\sqrt{\mathrm{ln}(\alpha_{\rm max}/\alpha(R,0))}}^{+\infty}\mathrm{exp}\left(S_z(R,0)\left(1-\frac{\alpha(R,0)}{\alpha_{\rm max}}e^{y^2}\right)-y^2\right)\mathrm{d}y
\end{equation}
\begin{equation}
B=\int_{\sqrt{\mathrm{ln}(\alpha_{\rm max}/\alpha(R,0))}}^{+\infty}\mathrm{exp}\left(S_z(R,0)\left(1-\frac{\alpha(R,0)}{\alpha_{\rm max}}e^{y^2}\right)\right)\mathrm{d}y
\end{equation}
\begin{equation}
C=\int_{\sqrt{\mathrm{ln}(\alpha_{\rm max}/\alpha(R,0))}}^{+\infty}y^2\mathrm{exp}\left(S_z(R,0)\left(1-\frac{\alpha(R,0)}{\alpha_{\rm max}}e^{y^2}\right)\right)\mathrm{d}y
\end{equation}

\begin{eqnarray}
\frac{\partial}{\partial R}\mathrm{ln}f(R,0)  =f(R,0)S_z(R,0)\Bigg[\left(\frac{\partial\mathrm{ln}\Sigma}{\partial R}+\frac{\partial\mathrm{ln}\alpha(R,0)}{\partial R}\right)\nonumber\\
 \Bigg(\frac{\sqrt{\mathrm{ln}\left(\alpha_{\rm max}/\alpha(R,0)\right)}}{\sqrt{\pi}\left(S_z(R,0)+1\right)}\left(\frac{\alpha(R,0)}{\alpha_{\rm max}}\right)^{S_z(R,0)+1}\nonumber\\
 -\frac{\mathrm{erf}\left(\sqrt{\left(S_z(R,0)+1\right)\mathrm{ln}\left(\alpha(R,0)/\alpha_{\rm max}\right)}\right)}{2\left(S_z(R,0)+1\right)^{3/2}}\nonumber\\
+\frac{2}{\sqrt{\pi}}\left(\frac{\alpha(R,0)}{\alpha_{\rm max}}\right)^{S_z(R,0)}\mathrm{ln}\left(\frac{\alpha(R,0)}{\alpha_{\rm max}}\right)\Bigg)\nonumber\\
+\frac{\partial\mathrm{ln}\Sigma}{\partial R}\frac{2}{\sqrt{\pi}}\left(\frac{\alpha(R,0)}{\alpha_{\rm max}}\right)^{S_z(R,0)}\left(A-B\frac{\alpha(R,0)}{\alpha_{\rm max}}\right)\Bigg]
\end{eqnarray}

\end{appendix}

\begin{acknowledgements}
The author thanks Matthieu Gounelle and S\'{e}bastien Fromang for interesting discussions on the topic of meridional circulation, as well as an anonymous referee whose comments helped in particular to interpret the qualitative difference between the two flows.
\end{acknowledgements}

\bibliographystyle{aa}
\bibliography{bibliography}

\end{document}